\documentclass[12pt]{iopart}
\usepackage{epsfig}

\begin{document}

\title{Optical Properties of Graphene-like Two Dimensional Silicene}

\author{Kamal Chinnathambi}
\address{ Indus Synchrotrons Utilization Division, Raja Ramanna Centre for Advanced Technology, Indore 452013, India \\
}
\ead{ckamal@rrcat.gov.in}
\author{Aparna Chakrabarti}
\address{ Indus Synchrotrons Utilization Division, Raja Ramanna Centre for Advanced Technology, Indore 452013, India \\
}
\author{Arup Banerjee}
\address{BARC Training School, Raja Ramanna Centre for Advanced Technology, Indore 452013, India\\
}
\author{S. K. Deb}
\address{ Indus Synchrotrons Utilization Division, Raja Ramanna Centre for Advanced Technology, Indore 452013, India \\
}

\begin{abstract}
We study optical properties of two dimensional silicene using density functional theory based calculations.  Our results on  optical response property calculations show that they strongly depend on direction of polarization of light, hence the optical absorption spectra  are different for light polarized parallel and perpendicular to plane of silicence. The optical absorption spectra of silicene possess two major peaks: (i) a sharp peak at 1.74 eV due to transition from $\pi$ to $\pi^*$ states and (ii) a broad peak in range of 4-10 eV due to excitation of $\sigma$ states to conduction bands. We also investigate the effect of external influences such as (a) transverse static electric field and (b) doping of hydrogen atoms (hydrogenation) on optical properties of silicene.  Firstly, with electric field, it is observed that band gap can be opened up  in silicene at Fermi level by breaking the inversion symmetry. We see appreciable changes in optical absorption due to band gap opening. Secondly, hydrogenation in silicene strongly modifies the hybridization and our geometry analysis indicates that the hybridization in silicene goes from mixture of sp$^2$ $+$ sp$^3$ to purely sp$^3$. Therefore, there is no $\pi$ electron present in the system. Consequently, the electronic structure and optical absorption spectra of silicene get modified and it undergoes a transition from semi-metal to semiconductor due to hydrogenation. 
\end{abstract}

\maketitle
\section{Introduction}
There has been a lot of interest in Graphene-like structure of silicon: Silicene, since it shows properties similar to those of graphene\cite{grap1,grap3,grap4,grap5}. The theoretical studies on silicene show that the charge carriers in this two-dimensional material behave like massless Dirac-Fermions due to the presence of linear dispersion around Fermi energy at a symmetry point K in the reciprocal lattice\cite{sili-ciraci1,sili-ciraci2}. Similar to its carbon counterpart - graphene, silicene is a potential candidate for applications in nanotechnology. The silicon based nanostructures have an important advantage that they are compatible with the existing semiconductor technology. Therefore, silicene and silicon nanoribbon have received much attention from both experimentalists and theoreticians\cite{sili-ciraci1,sili-ciraci2,sili-grown, sili-expt, lay, sinr1,sinr2,sili2,sili3,refpap1,refpap2,refpap3,refpap5}. Recently, silicene has been epitaxially grown on a close-packed silver surface Ag(111)\cite{sili-grown} and similarities of properties of silicene with graphene have been experimentally observed by Vogt \textit{et al} \cite{lay}. Though graphene possesses many novel properties, its application in nanoelectronic devices is limited due to its zero band gap and hence it is difficult to control the electrical conductivity of graphene. It is desirable to have band gap in materials in addition to their novel properties. Albeit band gap in graphene may be introduced by chemical doping but it is uncontrollable and incompatible with device applications. There exist a few studies on silicene which show that band gap in silicene can be opened up as well as varied over a wide range by applying transverse external static electric field\cite{ck-arxive, ni, drummond, ezawa}. 

Though there exist a few theoretical studies on geometric, electronic and vibrational properties of silicene,  no detailed study on optical properties of silicene is available in the literature. But studying optical properties of silicene is  important from both the fundamental as well as application point of view. Hence, in this work, we devote our study on optical properties of silicene using the state-of-art density functional theory (DFT) based calculations. Further, it is also interesting to explore how these properties could be altered due to external influences like external fields or chemical doping. Here, we choose two external influences namely (a) static transverse electric field and (b) hydrogenation - simplest possible doping, for investigating their effect on geometric, electronic and  optical properties of silicene. 
 
In the next section, we briefly outline the computational methods employed in the present work. The results and discussions are presented in section 3 and then followed by conclusion in section 4.   

\section{Computational details}
We use SIESTA package\cite{siesta1,siesta2,siesta3} for performing a fully self-consistent density functional theory (DFT) calculation by solving the standard Kohn-Sham (KS) equations. The KS orbitals are expanded using a linear combination of pseudoatomic orbitals proposed by Sankey and Niklewski\cite{sankey}. All our calculations have been carried out by using triple-zeta basis set with polarization function. The standard norm  conserving Troullier-Martins pseudopotentials\cite{tm} are utilized. For exchange-correlation potential generalized gradient approximation given by Perdew-Burke-Ernzerhof\cite{pbe} has been used. A cutoff of 400 Ry is used for the grid integration to represent the charge density and the mesh of k-points for Brillouin zone integrations is chosen to be 45$\times$45$\times$1. We used  finer mesh of 250$\times$250$\times$1 k-points for optical calculations. The convergence criteria for energy in SCF cycles are chosen to be 10$^{-6}$ eV. The structures are optimized by minimizing the forces on individual atoms (below 10$^{-2}$ eV/ $\AA$).  Super-cell geometry with a vacuum of 14 $\AA$ in the direction perpendicular to the sheet of silicene is used so that the interaction between adjacent layers is negligible.

\section{Results and Discussion}
As mentioned in the introduction, our main aim of the present work is to study optical properties of silicene. However, the optical property calculations within the framework of DFT, as implemented in SIESTA\cite{siesta1,siesta2,siesta3}, require eigen functions and eigen values of KS states. Therefore, before going into detailed discussion on results of optical properties, we first present our results on geometric and electronic structures of silicene which will be used later in calculations of optical properties. We also compare our results on geometric and electronic structure of silicene with those results available in literature. In next subsection, we start a discussion on results of geometric structure of silicene which is followed by discussions on electronic structures and optical properties respectively in subsection 3.2 and 3.3. 

\subsection{Geometric Structures} 

The optimized geometry of silicene is shown in Fig. 1 (a) and (b). The unitcell has two Si atoms (A and B) and the space group is P3m1. We observe that the minimum energy structure of the silicene is low-buckled with the lattice constant of 3.903 $\AA$. The bond length and bond angles between the silicon atoms are 2.309 $\AA$ and 115.4$^\circ$ respectively. The value of  bond angle in silicene lies in between those of sp$^2$ (120$^\circ$) and sp$^3$ (109.5$^\circ$) hybridized structures. 
This clearly shows that the hybridization in silicene is not purely of sp$^2$ type but a mixture of sp$^2$ and sp$^3$ types. The buckling in silicene is due to the weak $\pi$ - $\pi$ bond that exists between Si atoms since the Si-Si distance is much larger as compared to that in graphene (C-C =1.42 $\AA$).  The binding energy of the system increases due to buckling by increasing the overlap between $\pi$ and $\sigma$ orbitals. The presence of buckling may also be explained by Jahn-Teller distortion. Our results on geometry match well with the previous calculations\cite{sili-ciraci1,sili-ciraci2,sili2,sili3,refpap1}. Here, we quantify the amount of buckling in terms of buckling length $d$ which is defined as the vertical distance between atoms at sites A and B in the unitcell. The value of $d$ in silicene is 0.5 $\AA$. In case of planar structure, as that of graphene, the value of $d$ is zero. The importance of finite buckling in determining the electronic structure of silicene under the influence of external electric field will be described later.
\begin{figure}[ht]
\begin{center}
\includegraphics[width=8.5cm]{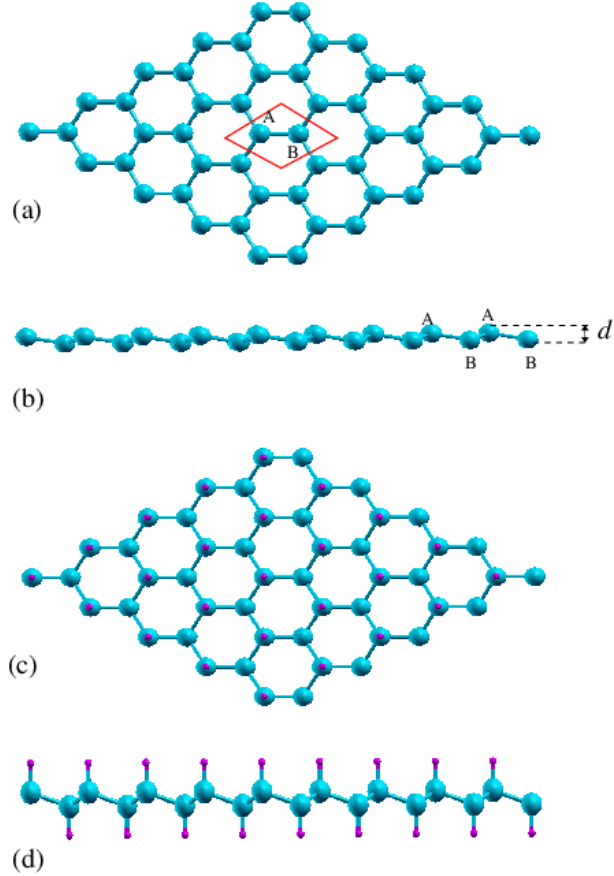}
 \caption{ The optimized geometry structures. Silicene : (a) Top and (b) side view of 5$\times$5 supercell. The lines in (a) represents the unitcell. The vertical distance between two Si atoms at sites A and B is represented by '$d$'. Hydrogenated Silicene : (c) Top and (d) side view of 5$\times$5 supercell. The small circles in (c) and (d) represent hydrogen atoms.}
\end{center}
 \end{figure}

We also optimize the geometry of silicene under two external influences such as transverse electric field and hydrogenation. We find that there is no appreciable change in the structure of silicene due to external electric field. In case of hydrogenation, we observe following modifications in the results of geometric structure. The hydrogenated silicene is called Silicane in analogy to the Graphane. The optimized structure of silicane (Si$_2$H$_2$) is shown in Fig. 1 (c) and (d). The value of lattice constant is 3.92 $\AA$.  The values of bond length between Si-Si and Si-H are respectively found to be 2.37 $\AA$ and 1.52 $\AA$. Further, the bond angles obtained are 111.1 $^\circ$ and  107.8$^\circ$ respectively for Si-Si-Si and H-Si-Si. The buckling length in silicane is 0.74 $\AA$ which is about 48 $\% $ more as compared to that of pure silicene. Our results on geometric properties of silicane match well with those results available in literature\cite{sili-hyd1,sili-hyd2,sili-hyd3,sili-hyd4}. It is interesting to note that the bond length and angle between Si atoms are respectively increased and decreased due to hydrogenation. Further, both these values are much closer to those of Si bulk (2.35 $\AA$ and 109.5$^\circ$). These results suggest that the hybridization in silicane is  sp$^3$-like.  The effect of changes in hybridization on electronic properties will be discussed later in next subsection. 

\subsection{Electronic Structures}
\begin{figure}[ht]
\begin{center}
\includegraphics[width=8.5cm]{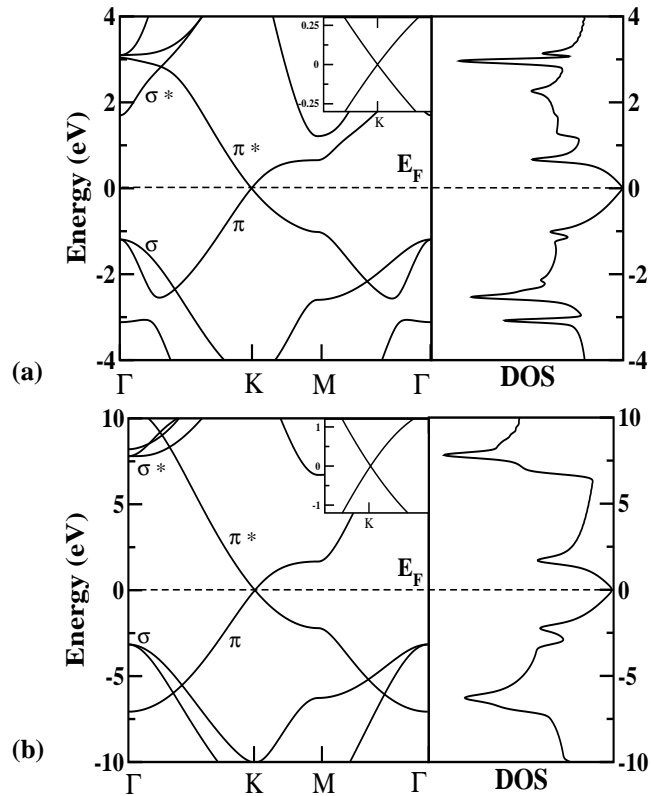}
 \caption{ Band  structure and density of states for optimized structures of  (a) silicene and (b) graphene. The energy of bands are with respect to Fermi level.}
\end{center}
 \end{figure}

The band structure along high symmetry points in Brillouin zone and density of states (DOS) of silicene and graphene are respectively shown in Fig. 2 (a) and (b). The figure (a) clearly indicates the semi-metallic behavior of silicene since the value of DOS at E$_F$ is zero and the conduction and valence band touch each other only at the high symmetry K point. The energy levels and the contribution of DOS just below and above Fermi levels are mainly due to $\pi$ and $\pi^*$ orbitals. The dispersion around K point near the Fermi level is linear (see insert in Fig. 2 (a)). The point in E-k diagram where the conduction and valence band touch each other at E$_F$ is called the Dirac point. Comparison of these results obtained for silicene (as presented in Fig. 2(a)) with those of graphene (in Fig. 2(b)) indicates that the electronic structure of silicene around E$_F$ is very similar to that of graphene. However, it is interesting to note that there is a difference in character of DOS between silicene and graphene well below  E$_F$. There exists a peak around -7 eV in graphene which is mainly due to $\sigma$ state but the corrponding peak in silicene is split due to mixing of $\pi$ and $\sigma$ states (or mixture of sp$^2$and sp$^3$ hybridization).  The presence of linear dispersion indicates that the charge carriers near Dirac point behave like massless Dirac Fermions since the dynamics of these carriers obeys relativistic Dirac-like equation. The relativistic Dirac-like Hamiltonian which describes the electronic structure of silicene around the Dirac point, similar to that of graphene\cite{oosting,wallace}, can be approximated as
\begin{eqnarray}
 \hat{H}= \left( {
\begin{array}{cc}  
\Delta & \hbar v_F (k_x-ik_y)  \\  
\hbar v_F (k_x+ik_y) & -\Delta  \\  
\end{array} } 
\right)
\end{eqnarray}
where $k$ and $v_F$ are momentum and Fermi velocity of charge carriers near Dirac point. The quantity $\Delta$ is the onsite energy difference between the Si atoms at sites A and B.  Due to the presence of inversion symmetry, the onsite energy difference $\Delta$ becomes zero, which leads to the linear dispersion around the Dirac point linear i.e. $E=\pm \hbar v_F k$.  Until now, the treatment for description of electronic structures of silicene is exactly similar to that of graphene. In the above mentioned discussion, the effect of spin-orbit coupling (SOC) is not included and very small gap of the order of $\mu$eV in graphene and about 1.55 meV in silicene may be opened up due to SOC\cite{grap-soc,sili-soc}. The band gap in graphene-like structure can be opened up if one can break the inversion symmetry and hence the value of  $\Delta$ becomes finite. Then, the dispersion around the Dirac point becomes
\begin{eqnarray}
 E=\pm \sqrt{\Delta^2+ (\hbar v_F k)^2}
\end{eqnarray}
Hence, in this case, the value of band gap opened is twice that of the onsite energy difference. 

\begin{figure}[]
 \begin{center}
 \includegraphics[width=8.5cm]{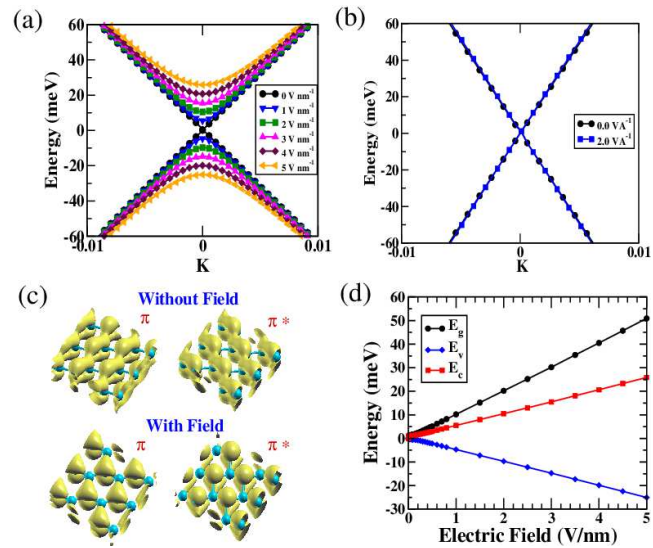}
 \caption{ The band structure of (a) Silicene and (b) Graphene around Dirac point for different strengths of transverse electric field. (c) Charge density distribution of $\pi$ and $\pi^*$ without and with electric field 5 V nm$^{-1}$. (d) The variation of band gap (E$_g$), valence (E$_v$) and conduction (E$_c$) band edges of silicene with the strength of transverse external field.}
 \end{center}
  \end{figure}

Now, we discuss the effect of transverse static electric field on the electronic structure of both silicene and graphene. The calculated band structures of silicene and graphene around Dirac point for various strengths of external electric field are shown in Fig. 3 (a) and (b). We observe from Fig. 3 (a) that a band gap can be opened up at Fermi level by external electric field due to the breaking of inversion symmetry. This is due to the fact that the potential seen by the Silicon atoms at the sites A and B is different. The presence of buckling in geometric structure of silicene plays an important role in breaking the inversion symmetry.  However, there is no gap (see Fig. 3(b) in graphene since potential seen by the carbon atoms at the sites A and B are same ($\Delta = 0$). Breaking of inversion symmetry can be clearly seen from modifications in charge density distribution of $\pi$ and $\pi^*$ states due to external electric field. In Fig. 3 (c), we plot the charge density distribution of these states, which lie at the symmetry K point,  with and without electric field. 
In the absence of external electric field, both the $\pi$ and $\pi^*$ states have same energy and also possess inversion symmetry. However, application of an external electric field makes the spatial distribution of charges above and below the silicence sheet to be different and also breaks the inversion symmetry. This leads to an opening up of the band gap in silicene. We also show that the band gap varies linearly with the strength of external electric field. These results are consistent with those of previous studies available in the literature\cite{ni,drummond, ezawa}. We find that the rate of increase in gap with strength of electric field is 0.1014 e$\AA$ which lies in between those estimated by Drummond \textit{et al}., (0.0742 e$\AA$)\cite{drummond} and Ni \textit{et al}., (0.157 e$\AA$)\cite{ni}. Moreover, the value of band gap can be tuned over a wide range. This result leads to an important advantage of silicene over graphene since there is no buckling in the latter and hence it is not possible to open up a gap by applying external electric field. We also observed that both valence and conduction band edges move symmetrically away from Fermi level when the electric field is applied. 
Further detail on calculations and results for effect of electric field on electronic structure of silicene can be found in the literature\cite{ck-arxive, ni,drummond, ezawa}.

\begin{figure}[ht]
\begin{center}
\includegraphics[width=8.5cm]{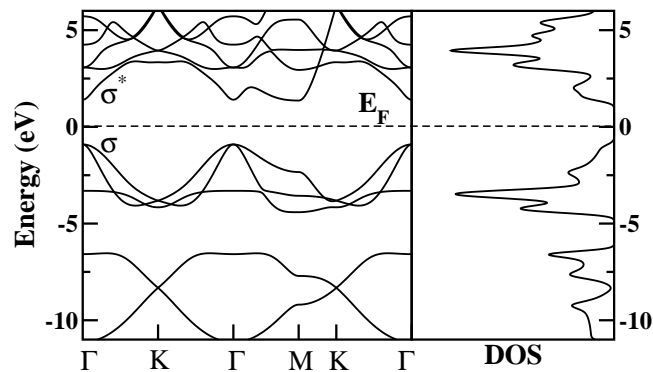}
\caption{ The band structure and DOS of hydrogenated Silicene.}
\end{center}
 \end{figure}

To understand the effect of hydrogenation on electronic properties of silicene, we calculate the electronic structure of silicane as given in Fig. 1 (c) and (d). The band structure and DOS of silicane is given Fig. 4.  We observe from this figure that electronic character of silicene changes from semi-metal to semiconductor due to hydrogenation. It is an indirect semiconductor with a band gap of 2.27 eV along $\Gamma$-M direction. However, the direct band gap of 2.31 eV at $\Gamma$ point is about 1.8\% higher. Our results on band structure match well with those available in the literature\cite{sili-hyd1,sili-hyd2,sili-hyd3,sili-hyd4}. For example, the value of band gap obtained in present work match well with about 2 eV obtained from LDA\cite{sili-hyd1, sili-hyd2}. However, it important to note that the values of band gap estimated from DFT calculations are underestimated by approximately about 30-50\% and hence the true gap may be larger. As mentioned earlier in geometry analysis, the hybridization in silicene is strongly modified due to hydrogenation and it becomes  sp$^3$-like.  The semi-metallic character of undoped silicene is due to presence of $\pi$ orbitals near $E_F$ which contains the delocalized $\pi$ electrons. However,  due to hydrogenation, the weak $\pi$ bond present in undoped silicene is easily broken and is replaced by strong covalent $\sigma$ bond between Si and hydrogen atoms. 
It is clearly seen from Fig. 4 that $\pi$ band which crosses $E_F$ in undoped silicene has disappeared and a new $\sigma$ band corresponding to Si-H bond has appeared in band structure of silicane. So, there is no $\pi$ electron present in system and hence there is a transition from semi-metallic to semiconductor character. It is interesting to note that the hydrogenation only modifies the electronic states which lie near $E_F$ and the features of $\sigma$ bands of undoped silicene remains similar. These results again suggest that behavior of silicene is akin to graphene. 

\subsection{Optical properties}

Having discussed the results on geometric and electronic structures of silicene in previous subsections, we now focus our attention on discussion on  results obtained for optical response properties of silicene. The optical response properties of silicene are calculated by employing first order time-dependent perturbation theory as implemented in SIESTA package\cite{siesta1,siesta2,siesta3}. For accurate calculation of optical response properties, it is necessary to use finer mesh of k-points and hence we used large mesh size of 250$\times$250$\times$1 in the present calculations.  In optical property calculations, the imaginary part of dielectric functions for the light polarized parallel and perpendicular to the plane of silicene sheet are evaluated. The real part of dielectric functions is then obtained via Kramers-Kronig relation.  It is clearly seen from Fig. 5 (a) that dielectric function of silicene strongly depends on the direction of polarization of light since parallel ($\varepsilon_x$ and $\varepsilon_y$ represented by $\varepsilon_\parallel$)  and perpendicular ($\varepsilon_z$ represented by $\varepsilon_\perp$) components of $\varepsilon$ are different. It is also observed from the above figure that parallel and perpendicular  components respectively dominate lower (below approximately 5 eV) and higher (above  5 eV) energy regime. The presence of anisotropy in dielectric function is a consequence of two dimensional nature of the silicene sheet. We also calculate the total response of $\varepsilon_T$ of silicene : bare (bottom panel in Fig. 5 (a)), with electric field of 1.0 V$\AA^{-1}$ (Fig. 5 (b)) and hydrogenated (Fig. 5 (c)). These resuls have been used to calculate the optical absorption spectra of silicene. 
\begin{figure}[ht]
\begin{center}
\includegraphics[width=8.5cm]{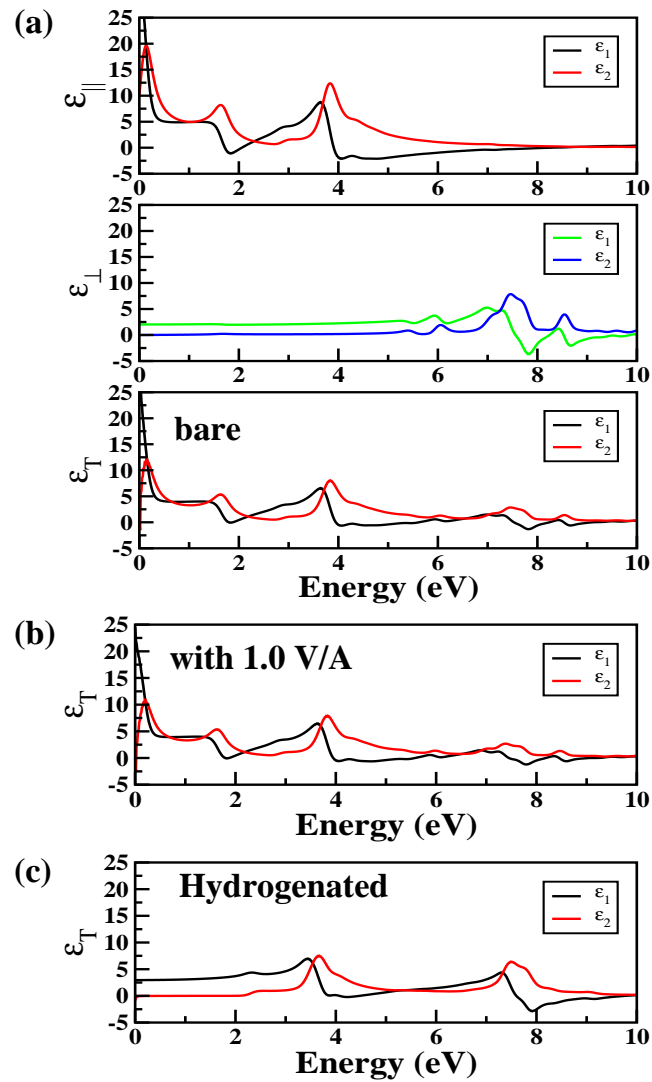}
\caption{(a) The calculated real ($\varepsilon_1$) and imaginary ($\varepsilon_2$) part of dielectric function of silicene: parallel (top), perpendicular (middle) components and total (bottom). (b) $\varepsilon_T$ for silicene with electric field (1.0 V$\AA^{-1}$). (c) $\varepsilon_T$ for hydrogenated silicene  }
\end{center}
\end{figure}

The calculated optical absorption spectrum of silicene is given in Fig. 6(a). The absence of cut-off energy in absorption coefficient suggests that there  is no band gap in silicene. The analysis of optical absorption spectra shows that silicene possesses two major peaks in energy range from 0 to 6 eV. These two peaks correspond to the two important transitions that occur between electronic states of silicene. They are (a) smaller peak around 1.74 eV and (ii) intense broader peaks, with maximum at 3.94 eV, which extend beyond 6 eV. First peak corresponds to the transition from states $\pi$ to $\pi^*$ which are close to the Fermi level. Hence the lower energy spectra of silicene are mainly dominated by $\pi$ electron present in these states. 
The value of the absorption peak matches well with 1.69 eV of peak-to-peak (above and below $E_F$) energy difference in DOS (see in Fig. 2(a)).  It is interesting to note that shape of this peak is symmetric which is due to symmetric dispersion of $\pi$ and $\pi^*$ bands respectively below and above $E_F$.  The broad second peak corresponds to transition from occupied $\sigma$ to unoccupied $\sigma^*$ state in the conduction bands. 
The broad energy range of this transition is due to the large band width of both $\sigma$ and $\sigma^*$ states. 

\subsubsection{Effect of External Electric Field}

\begin{figure}[ht]
\begin{center}
\includegraphics[width=8.5cm]{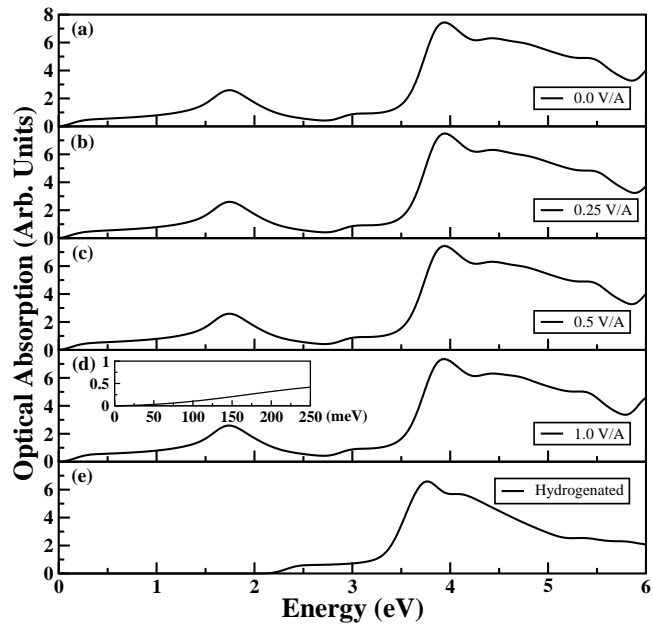}
\caption{The optical absorption spectra of Silicene: (a) without electric field, (b) with 0.25 V$\AA^{-1}$, (c) with 0.5 V$\AA^{-1}$,(d) with 1.0 V$\AA^{-1}$ and (d) hydrogenated. }
\end{center}
 \end{figure}

The optical absorption spectra of silicene when it is subjected to different strength of transverse static electric field are given in Fig. 6(b)-(d). The figures (See insert in (d)) clearly indicate that there is no absorption of light having energy below a certain cut-off value. The value of cut-off also increases with the strength of electric field. These observations suggest that there is a band gap opening in silicene due to external electric field and also band gap increases with increasing strength of electric field. These results are also consitent with our discussions on effect of transverse electric field on electronic structure of silicene which are presented in previous subsection (Fig. 3 (a) and (d)). Interestingly, the features of spectra with electric field, such as the width and position of peaks, corresponding to different transitions remain similar to those of silicene without electric field. We infer from these observations that main effect of electric field is to influence the electronic states of silicene which are close to Fermi level. 

\subsubsection{Effect of Hydrogenation}
The calculated optical absorption spectra of silicane is given in Fig. 6(e). It is clearly seen that there is transition from semi-metallic to semiconductor behavior since there is no absorption upto a threshold value of band gap. The main influence of hydrogenation on optical absorption spectra is the disappearance of first peak around 1.74 eV since, as discussed in previous section, there are no  $\pi$ bands present in system. Hence, no transition between $\pi$ and $\pi^*$ can occur.  However, the second peak which correspond to $\sigma$-to-$\sigma^*$ transition is not much influenced by hydrogenation as these states are not modified by this effect.  This observation again suggest that main effect of hydrogenation is to modify the  $\pi$ and  $\pi^*$ bands which are close to $E_F$.

\section{Conclusion}
In summary,  we have carried out abinitio DFT calculations to study optical response properties of silicene. Our results on geometric and electronic structures of silicene match well with the results available in literature. We observed from our results on optical response property calculations that the dielectric function strongly depend on direction of polarization of incident light. The anisotropic response in dielectric function is a consequence of two dimensional characteristic of silicene. The optical absorption spectra of silicene possess two major peaks due to $\pi$-to-$\pi^*$ and $\sigma$-to-$\sigma^*$ transitions. In this work, we also studied the effect of (a) transverse static electric field and (b) hydrogen doping on geometric, electronic and optical properties of silicene. We have shown that band gap in silicene can be opened up near Fermi level by applying external electric field. We observed appreciable changes in optical absorption due to band gap opening. In case of hydrogenation, we observed from geometry analysis that the hybridization in silicene goes from mixture sp$^2$ $+$ sp$^3$ to sp$^3$. Consequently, the electronic structure of silicene gets modified drastically and it undergoes a transition from semi-metal to insulator due to hydrogenation. Therefore, we infer from our results that both these external influences strongly modify the electronic properties of silicene mainly the states ($\pi$ and $\pi^*$) which are close to $E_F$. The results obtained in the present can be directly verifiable from experimental studies like transport properties and optical measurements.

\section{Acknowledgments}
Authors thank Dr. P. D. Gupta for encouragement and support. Thanks also to Dr. J. Jayabalan for critical reading of the manuscript. The support and help of Mr. P. Thander and the scientific computing group, Computer Centre, RRCAT is acknowledged.

\end{document}